\documentclass[reprint,aps,prl,amsmath,amsfonts,sort&compress,superscriptaddress,noeprint,amssymb,floatfix]{revtex4-2}

\usepackage{times}
\usepackage{epsfig}
\usepackage{graphicx}
\usepackage{epstopdf}
\usepackage{bm}
\usepackage{color}
\usepackage[english]{babel}
\usepackage{microtype}
\usepackage[breaklinks=true,colorlinks=true,citecolor=blue,urlcolor=blue,linkcolor=blue,pdfencoding=auto,psdextra]{hyperref}

\usepackage{xcolor}
\usepackage{soul}

\newcommand{\s}{\,s$^{-1}$}
\newcommand{\ergs}{\mbox{\,erg\,\s}}

\newcommand{\lbol}{L_{\rm bol}}
\newcommand{\lx}{L_{\rm x}}

\newcommand{\lc}{l_{\rm c}}
\newcommand{\rl}{\mathcal{R}_L}
\newcommand{\rc}{r_{c}}

\newcommand{\rostom}[1]{{\bf \color{blue} #1}}

\def\app#1#2{%
  \mathrel{%
    \setbox0=\hbox{$#1\sim$}%
    \setbox2=\hbox{%
      \rlap{\hbox{$#1\propto$}}%
      \lower1.1\ht0\box0%
    }%
    \raise0.25\ht2\box2%
  }%
}

\usepackage{float}
\definecolor{LightCyan}{rgb}{0.88,1,1}

\newcommand{\aap}{A\&A}

\newcommand{\apjl}{ApjL}
\newcommand{\apjs}{ApjS}

\usepackage{xr}
\begin{document}

\title{The Interplay between accelerated Protons, X-rays and Neutrinos in the Corona of NGC~1068: Constraints from Kinetic Plasma Simulations}% Force line breaks with \\

\author{Rostom Mbarek}
\altaffiliation{Neil Gehrels Fellow}
\email{rmbarek@umd.edu}
\affiliation{Joint Space-Science Institute, University of Maryland, College Park, MD, USA}%
\affiliation{Department of Astronomy, University of Maryland, College Park, MD, USA}
\affiliation{Astrophysics Science Division, NASA Goddard Space Flight Center, Greenbelt, MD, USA}

\author{Alexander Philippov}%
%\email{sashaph@umd.edu}
\affiliation{Department of Physics, University of Maryland, College Park, MD, USA}%
\affiliation{Institute for Research in Electronics and Applied Physics, University of Maryland, College Park, MD 20742, USA}%

\author{Alexander Chernoglazov}%
%\email{achernog@umd.edu}
\affiliation{Department of Physics, University of Maryland, College Park, MD, USA}%
\affiliation{Institute for Research in Electronics and Applied Physics, University of Maryland, College Park, MD 20742, USA}%

\author{Amir Levinson}%
%\email{levinson@tauex.tau.ac.il}
\affiliation{The Raymond and Beverly Sackler School of Physics and Astronomy, Tel Aviv University, Israel}

\author{Richard Mushotzky}%
%\email{rmushotz@umd.edu}
\affiliation{Joint Space-Science Institute, University of Maryland, College Park, MD, USA}
\affiliation{Department of Astronomy, University of Maryland, College Park, MD, USA}

\begin{abstract}
We examine properties of accelerated protons potentially responsible for the neutrino excess observed in the direction of NGC~1068, using constraints from kinetic Particle-in-Cell (PIC) simulations. We find that \emph{\textbf{i})} coronal X-rays and Optical/Ultra-Violet light in the inner disk lead to efficient absorption of hadronic $\gamma$-rays within 100 Schwarzschild radii from the black hole; \emph{\textbf{ii})} protons accelerated from the coronal thermal pool cannot account for the observed neutrinos; and \emph{\textbf{iii})} explaining the observed signal requires an injection of protons with a hard spectrum, peaking at $\gamma_p\sim 10^3-10^4$, into the turbulent magnetically-dominated corona, where they are confined and re-accelerated. The resulting neutrino signal can be consistent with IceCube observations. In our most favorable scenario, the injected protons are pre-accelerated in intermittent current sheets in the vicinity of the black hole, occurring either at the boundary between the disk and the outflow or during magnetic flux eruption events.
\end{abstract}

\date{\today}

\maketitle

\paragraph{\bf Introduction.}

Recent observations of NGC~1068 show a significant neutrino flux \citep{IceCube-NGC1068} with a luminosity $L_\nu \lesssim 10^{42}$erg/s that dwarfs its expected $\gamma$-ray counterpart \citep{aartsen+20,IceCube-NGC1068}.
It has hence been suggested that opaque Active Galactic Nucleus (AGN) cores can be significant high-energy neutrino sources, where dense radiation attenuates $\gamma$-rays while contributing to neutrino emission \citep[e.g.][]{murase+20,inoue+20,kheirandish+21,murase22,halzen+22,kurahashi+22,eichmann+22,halzen23,fang+23,hooper+23,fiorillo+23b}.
Specifically, highly-magnetized coronae are promising particle accelerators that can also produce neutrinos mainly through interactions with coronal X-rays \citep[e.g.][]{stecker+91,inoue+19,murase22}, within a region $\lesssim 100 r_g$ \footnote{$r_g= 2GM_{\rm BH}/c^2$, where $M_{\rm BH}$ is the black hole (BH) mass} \citep{murase22}.
Nonetheless, a framework that self-consistently explains these observations is lacking as the complex interplay of particle acceleration, disk properties, and radiation fields is decisive. In this Letter, we investigate whether magnetized coronae in the vicinity of supermassive black holes (BHs) are viable candidates to be neutrino emitters.

Initially, we study the effects of radiation and magnetic fields in coronae relying on observations of X-rays and Optical/UV (OUV) radiation of AGNs.
%Variability analyses of microlensed X-rays in quasars show that radiation
X-rays in the nuclear regions emanate from compact coronal regions \citep{fabian12,miller+13} with half-light radii $\sim 6 r_g$ \citep{fabian+15}, and could lie within $\lesssim 10 r_g$ above the BH \citep[e.g.][]{fabian+09,dai+10,demarco+11,kara+13,uttley+14,wilkins+21}.
These X-rays from the central parts of AGNs are produced by the comptonization of disk photons on electrons energized in the coronae \citep[e.g.][]{katz76,pozdnyakov+77}. 
We rely on a recent understanding of the dissipation of magnetic energy and its conversion to radiation in collisionless magnetic reconnection and turbulence, in first-principles models of X-ray spectra of X-ray binaries \citep[e.g.][]{beloborodov17,groselj+24}. Similar conditions are expected in the coronae of NGC~1068, a Compton thick AGN with a thin disk where matter obscures nuclear X-ray emission because of a large column density. 
As for OUVs, they are found to be directly correlated with X-rays in Seyferts \citep[e.g.][]{lusso+10,lusso+12}. Henceforth, we construct a neutrino emission framework that accounts for particle acceleration and the radiation properties of coronal regions, concomitant with the effects of neutrino production mechanisms (proton-proton, $pp$, and photomeson, $p\gamma$) based on analytical arguments and kinetic particle-in-cell (PIC) plasma simulations.

\paragraph{\bf Parent Protons of Detected Neutrinos.}
From energy conservation, we write the neutrino flux $\phi_\nu$, based on the parent proton spectrum $\phi$ considering that the multi-pion production channel is dominant in the TeV-PeV range, %assuming that 1 neutrino is produced for $p\gamma$ and $pp$ interactions \citep{mbarek+23},
\begin{equation}\label{eq:nu-eq}
E_\nu^2 \phi_\nu (E_\nu) \approx
    \psi_{p \rm x} \kappa_{\nu} E_p^2 \phi (E_p) 
\end{equation}
where $E_p$ is the proton energy, and $E_\nu = \alpha E_p$ the neutrino energy, with $\alpha \simeq 1/20$. $\kappa_{\nu}$ is akin to an optical depth for neutrino production, such that $\kappa_{\nu}(E_p) = \text{min}(t^{-1}_\nu \langle t_{\rm esc} \rangle, 1)$, with $t_\nu$ the proton cooling time, and $t_{\rm esc}$ the proton escape time from the corona.
%Max$(\kappa_{\nu})=3$ is the number of interactions needed for a proton to lose 1/2 of its energy, as protons lose $E_p/5$ per $p\gamma$ interaction in this range.
$\psi_{p \rm x}$ is an interaction-dependent scaling where $\psi_{pp} = 1/2$ (from the charged to neutral pion ratio) and $\psi_{p\gamma} = 3/4 \psi_{pp}$ (adding a scaling from pion and muon decay).
The proton luminosity $L_p$ needed to explain the 1-10TeV neutrino spectrum is then,
\begin{equation}\label{eq:Lp}
	L_p \approx 4 \pi D^2 (1 + z)^2 \frac{E_\nu^2 \phi_\nu (E_\nu)}{\psi_{p \rm x} \kappa_{\nu}}  
\end{equation}
where $z$ is the redshift, $D$ is the distance to the source, and $E_\nu^2 \phi_\nu (E_\nu)$ is extracted from IceCube data \citep{IceCube-NGC1068}.

\begin{figure}
	\centering
	\includegraphics[width=0.46\textwidth,clip=false,trim= 0 0 0 0]{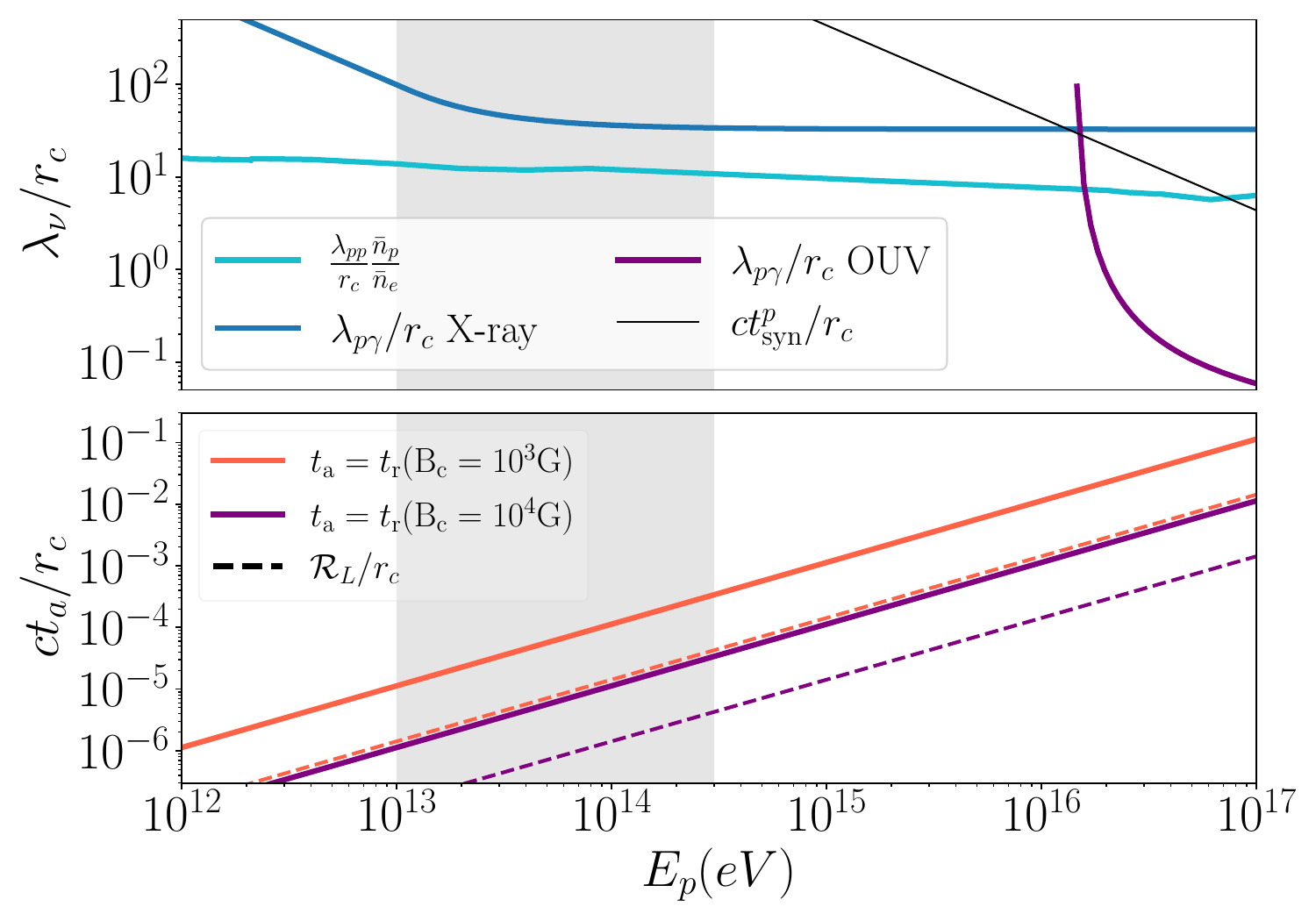}
	\caption{Upper Panel: Cooling length for proton synchrotron, $pp$, and $p\gamma$ interactions in the putative corona of NGC~1068, in an energy range encompassing the observed neutrino signal (the shaded region).
	Lower Panel: Acceleration distance, $ct_{\rm a}$,  for relativistic magnetic reconnection, $ct_{\rm r}(B_c)$, along with proton Larmor radii, $\rl$. As for the stochastic acceleration distance, $ct_{\rm s}(\sigma)/\rc \simeq 10^{-1}$ for $l_0 \sim r_g$ \citep{comisso+19}, with a relatively unknown energy dependence.}
	\label{fig:cooling-in}
\end{figure}

We retrieve the required slope of the distribution of accelerated protons, $\phi(E)\propto E^{-s}$, from Eq.~\ref{eq:nu-eq}, where the neutrino spectrum, $\phi_\nu \propto E^{-q} \propto E^{-s} \kappa_{\nu}(E)$, and $ q = 3.2 \pm 0.2$ \citep{IceCube-NGC1068}. As we discuss below, for realistic plasma conditions and properties of the radiation fields, $t_\nu $ is mostly energy-independent (Fig.~\ref{fig:cooling-in}), and $t_{\rm esc}$ scales weakly with $E_p$, $t_{\rm esc} \propto E_p^{-\delta_s}$, where $\delta_s \sim 0.3$. We find that protons are efficiently confined in the corona, resulting in $\kappa_{\nu}\approx {\rm const}$. Therefore, $s \in [3.0,3.4]$, which sets the coronal plasma magnetization $\sigma$, defined as the ratio of the magnetic energy density to the rest mass energy density, thanks to kinetic simulation results.

The observed neutrino to bolometric luminosity ratio, $L_\nu/\lbol \approx 10^{-3}$ \citep{woo+02,herrero+11}, implies that the accelerated protons' spectral luminosity, $L_p(E_p)$, peaks for $E_p\sim$ 1-10 TeV. 
Otherwise, if the inferred proton spectrum is extrapolated down to lower energies, $E_{p,\rm min}\lesssim 100$ GeV, the injected proton power would exceed the bolometric luminosity, viz., $L_p(E_{p,min})>\lbol\simeq 7 \times 10^{44}$erg/s \citep{bauer+15}, violating energy conservation. Consequently, proton acceleration from the coronal thermal pool cannot account 
for the neutrino signal, unless the coronal plasma magnetization $\sigma$ is unlikely large, $\sigma \gtrsim 10^3$.  The reason is that particle acceleration in magnetized plasma (by either turbulence or magnetic reconnection) tends to produce a soft spectrum ($s >2$) above the equipartition energy ($\sim \sigma$ in collisionless plasma) and a hard spectrum ($s\lesssim 1$) below it.
However, a low-density proton population pre-accelerated to $ \gamma_{p,\rm inj} \sim 10^3-10^4$, 
in a disparate region of large magnetization, $\sigma_p \gtrsim 10^3-10^4$, then injected and reaccelerated in the turbulent 
coronal plasma, can accommodate the energetic demands.  
In the scenario discussed below, proton pre-acceleration is supposed to occur in high-$\sigma_p$ intermittent current sheets, at either the outflow boundary or current sheets in BH magnetic flux eruptions.

\paragraph{\bf Coronal Radiation Fields.}
Considering their importance for $p\gamma$ interactions and $\gamma$-ray obscuration, we discuss radiation fields in the inner regions of NGC~1068.
We adopt the unified model of AGNs, where the apparent difference between the Seyfert I and II classes is solely due to orientation to the observer. We can then rely on type~I AGN (visible nucleus) for clues on radiation fields in type~II AGNs (e.g. NGC~1068). In this sense, we consider coronal X-ray emission, along with its corresponding Optical/UV (OUV) emission through the $\alpha_{\rm ox}$ relation as the dominant radiation fields (See Appendix).

As for $\gamma$-rays produced through $pp$ and $p\gamma$ processes, they are suppressed through the Breit-Wheeler process (See Appendix).
Generally speaking, 
%interactions with coronal X-rays are not sufficient to curb the TeV $\gamma$-ray emission.
%However, 
AGN OUV emission inferred from Eq.~\ref{eq:Lxuv} can attenuate $\gamma$-rays within $< 100 r_g$ to a level consistent with observational limits \citep{MAGIC19}, while GeV $\gamma$-rays are attenuated by coronal X-rays.
More data in the TeV range will put firmer constraints on consistency with the hadronic emission scenario. A future study of the ensuing $\gamma$-ray cascades can help to interpret Fermi data \citep{FERMI20}.

\paragraph{\bf Coronal Magnetic Fields and Composition.}
Considering its crucial role in particle acceleration and radiation production, we constrain the coronal magnetic field strength, $B_c$. If magnetic reconnection drives coronal X-ray production \citep[e.g.,][]{beloborodov17}, where the inverse Compton (IC) lepton cooling timescale is much shorter than the dynamical time, most of the magnetic energy dissipated in reconnection is quickly converted into radiation, such that $U_{\rm x} \sim \beta_r U_B =  \beta_r B_c^2/8\pi $, where $\beta_r= 0.1$ \citep[e.g.,][]{lyubarsky05--} is the dimensionless reconnection velocity. We find $B_c \approx 2 \times 10^4$G. Conversely, if magnetized, large-amplitude fluctuations are present, i.e. $\delta B\sim B_c$, turbulence is the main driver. We expect a balance between the energy carried away by escaping radiation, $\sim U_{\rm x}/t_{\rm \gamma,esc}$, and the turbulent cascade power, ${(\delta B)^2}/(4\pi t_0)$, where $t_{\rm \gamma,esc} = (\tau_{\rm T} + 1)l_{\rm esc}/c$ is the photon escape time associated with diffusion over scale $l_{\rm esc}$, $\tau_{\rm T}\gtrsim 1$ \footnote{The condition 4 $kT_e/(m_e c^2) \tau_{\rm T}^2 \sim 1$ should be satisfied, for thermal electrons cooled by seed photons through inverse Compton \citep[e.g.][]{rybicki+79,beloborodov17}. This places $\tau_{\rm T} \gtrsim 1$ for coronae exhibiting a hard state at $kT_e\sim 10-100$ keV. Observations of AGN infer similar optical depths \citep[e.g.][]{fabian+15}.} the Thomson optical depth, and $t_0$ the eddy turnover time at the
turbulence driving scale, $l_0$ \citep{groselj+24} \footnote{$l_{\rm esc}$ is defined as a diffusion scale as required by the balance between the energy carried away by escaping radiation and the turbulent cascade. Assuming that $l_{\rm esc}$ is also equal to the turbulence driving scale $l_0$ in a self-consistent environment $l_{\rm esc} \sim l_0$ should be a fair assumption.}. For $l_{\rm esc}\sim l_{\rm 0}$ and $\tau_{\rm T}\sim 1$, we find $B_c \approx 2 \times 10^3$G. Overall, we expect $B_c \sim 10^3$-$10^4$G, depending on energy dissipation, consistent with sub-equipartition field strength in thin disks \citep[e.g.,][]{beloborodov17}.

Plasma in BH coronae consists of electrons, protons, and most likely positrons, produced in two-photon collisions of compotonized photons. The existence of positrons is motivated by observations of coronae lying close to the pair balance line, where electron and positron densities are equal \citep[e.g.,][]{fabian+15}, but the lepton to proton total density ratio, $\bar{n}_e/\bar{n}_p$, is uncertain. $\bar{n}_e$ is self-regulated such that $\tau_{\rm T} \sim 1$, resulting in $\bar{n}_e\simeq {\tau_{\rm T}}/({\Sigma_T \rc}) \simeq 10^{11}{\rm cm}^{-3}$\footnote{Note that this value  exceeds the pair density expected from absorption of the pionic gamma rays alone.}. We can then calculate the pair magnetization parameter, $\sigma_{\pm}=B^2/(4\pi \bar{n}_e m_e c^2)=(2\ell/\tau_{\rm T}) (U_B/U_{\rm x})$, thus $\sigma_{\pm,{\rm turb}}\approx  \ell/\tau_{\rm T}$, for the turbulence scenario \footnote{For the reconnection scenario, $U_B/U_{\rm x}\approx 1/\beta_{\rm rec}$, and for a current sheet of the length $\approx r_c$, one can estimate $\tau_{\rm T}=\Sigma_T \bar{n}_e h\approx 1$, where $h\approx \beta_{\rm rec} r_c$ is the characteristic width of the current layer \citep{beloborodov17}, resulting in a similar estimate.}, where $\ell=\Sigma_T U_{\rm x} r_c/(m_e c^2)\approx 10$ is the radiative compactness parameter \footnote{We note that this value corresponds to the average radiation flux from the BH, and, thus, average magnetic field strength. Compactness near dissipation regions, which can be localized in time and space, could be significantly higher.}. Proton magnetization is then $\sigma_{\rm p}=\sigma_{\pm} (m_e/m_p) (\bar{n}_e/\bar{n}_p)$, and $\sigma_{\rm p}\lesssim 10$, for $\bar{n}_e/\bar{n}_p \lesssim m_p/m_e$. Below, we focus on the mildly relativistic scenario $\sigma_{\rm p}\sim 1$ as kinetic simulations strongly hint to this case for $s \approx 3$. %and discuss how results change for $\sigma_{\rm p}\lesssim 1$.
Note that these coronal protons are magnetically energized such that $\langle \gamma_p \rangle \sim \sigma_p$, and so $L_p \leq \lx$.

\paragraph{\bf Proton Escape Time.} Strong turbulent magnetic fields in the corona can lead to efficient confinement of accelerated protons. If magnetic fields are dynamically important in the vicinity of BHs, we estimate $r_g \lesssim \lc \lesssim h $, where $\lc$ is the coherence length of the B-field \footnote{The coherence length $\lc$ should in principle also scale with the turbulence driving scale $l_0$ in a self-consistent environment.} and $h\sim r_g$ is the scale height of the thin disk. This assumption constrains $t_{\rm esc}$, since for $B_c \sim 10^3-10^4$G, the proton Larmor radii are $\rl \ll \lc$, which eventually results in  significant scattering in the corona.
The proton mean free path is set as $\lambda_s \sim \lc (\rl/\lc)^{\delta_s} $ for $\rl/\lc \ll 1$,  where $\delta_s  \approx 0.3$ corresponds to scattering on intermittent small-scale field reversals as is likely appropriate for large-amplitude turbulence; the exact value of the exponent is a subject of active investigations \citep[e.g.,][]{lemoine23,kempski+23}. For strong diffusion, $\rl/\lc \ll 1$ so that particles random walk, and we get $\langle c t_{\rm esc} /r_c \rangle \sim r_c/\lambda_s \sim \left({r_c}/{l_c}\right) \left({ r_g}/{\rl}\right)^{1/3}$ \citep{effenberger+18}, and,
\begin{equation}\label{eq:tesc}
	\langle c t_{\rm esc} /r_c \rangle \sim 200 \left( \frac{10^{14} {\rm eV}}{E_{\rm p}} \cdot \frac{B_c}{10^{3}{\rm G}} \cdot \frac{M_{\rm BH}}{10^7M_{\odot}} \right)^{1/3}.
\end{equation}
It follows that if a mildly relativistic coronal inflow/outflow is present, with a velocity $\gtrsim 5 \times 10^{-3} c$, the proton residence time will be limited by the coronal inflow/outflow time.

\paragraph{\bf Proton Cooling in the Corona.}
\paragraph{ Photomeson ($p\gamma$) interactions.}
We calculate the $p\gamma$ mean free path $\lambda_{p \gamma}=c/t^{-1}_{p \gamma}$ based on the effects of X-rays and OUV using Eq.~\ref{eq:photomes} \citep{stecker68}. See details in the Appendix.

\paragraph{ Proton-proton ($pp$) interactions.}
The $pp$ mean free path is $\lambda_{pp} = (\bar{n}_p \Sigma_{pp})^{-1}$, where $\Sigma_{pp}$ is the cross section for $pp$ interactions \citep{PDG18} and $\bar{n}_p$ is the density of coronal protons. 
We can further express it as $\lambda_{pp}=(r_c/\tau_{\rm T}) (\Sigma_T/\Sigma_{pp})(\bar{n}_e/\bar{n}_p)$.

%$\sigma_\pm$
\paragraph{Cooling Synthesis.}

In the upper panel of Fig.~\ref{fig:cooling-in}, we show the normalized mean free path $\lambda/r_c$ for $pp$ and $p\gamma$ interactions, along with the effects of synchrotron cooling \footnote{See Supp. Material (SM) for a discussion of the synchrotron cooling on protons; in particular, Eq.~6
%\ref{SuppMeq:tsyn} 
in the SM was used to plot the limit in Fig.~\ref{fig:cooling-in}}. We also account for the potential impact of pion synchrotron cooling \citep[e.g.][]{guarini+23}. 
For $pp$ interactions, we plot $\lambda_{pp} (\bar{n}_p/\bar{n}_e)$, insensitive to $\bar{n}_e/\bar{n}_p$. 
We note \emph{i)} that protons invariably interact through $p\gamma$ in the corona because of confinement (Eq.~\ref{eq:tesc} and Fig.~4
%\ref{SuppMfig:tnu-tesc} 
in Supp. Material (SM)), and \emph{ii)} the presence of an OUV ``wall'' that sets a maximum energy $E_{\rm wall} \lesssim 10^{17}$eV, beyond which a) $t_a>\lambda_{p \gamma}/c$ and b) protons don't escape the corona before interacting with the 2500\AA~component. $pp$ interactions could be significant, for $\bar{n}_e/\bar{n}_p\lesssim 5$, which is equivalent  to $\sigma_p \lesssim 0.1$. For mildly relativistic magnetizations, $\sigma_p\sim 1$, corresponding to $\bar{n}_e/\bar{n}_p\sim 200$, $p\gamma$ interactions dominate proton cooling. In the following, we further constrain $\sigma_p$ and $\bar{n}_e/\bar{n}_p$. Moreover, the proton cooling time is shorter than the escape time due to diffusion (Eq.~\ref{eq:tesc} \footnote{See a separate Figure in the SM showing the ratio of the escape time and cooling times to different processes.}), resulting in efficient conversion of proton energy into neutrinos.

\paragraph{\bf Proton Acceleration in the Coronal plasma.}
In what follows, we probe proton acceleration in the corona using PIC plasma simulations. We consider the two most likely processes producing efficient particle acceleration in magnetically-dominated plasma: relativistic magnetic reconnection and turbulence. We verify whether features of reconnection-accelerated and/or turbulence-accelerated coronal protons are compatible with the IceCube signal. We conclude that protons accelerated from the thermal pool \emph{cannot} explain the IceCube results for both acceleration mechanisms. The most likely source of the observed neutrinos are low-density protons pre-energized in a region with a large $\sigma_p$ through reconnection, and later confined and re-accelerated in the turbulent corona. We provide constraints on the magnetization in the corona, $\sigma = {B^2}/{4\pi c^2(m_e \bar{n}_e+m_p \bar{n}_p)} =  \sigma_\pm/(1 +{\sigma_\pm}/{\sigma_{p}})$, and, thus, $\bar{n}_e/\bar{n}_p$, based on the observed neutrino flux and spectral slope, $s$, and properties of the particle acceleration mechanisms.

It has recently been shown that highest-energy particles in 3D relativistic reconnection are accelerated while on ``free-streaming'' trajectories, bouncing between the two converging upstream flows \citep{kowal+12,zhang+21b, chernoglazov+23}. The acceleration time of these particles, $t_{\rm r}$, corresponds to a distance $ct_{\rm r}=c/(\beta_{\rm r}\omega_B)$, where $\omega_B=eB_{\rm c}/(m_p c\gamma_p)$ is the gyrofrequency of accelerated protons. In magnetized turbulence, highest energy particles are accelerated by stochastic scattering off turbulent fluctuations \citep{comisso+19}, corresponding to an acceleration distance,  $ct_{\rm s} \sim {3 l_0}/{\sigma} $, where $l_0\sim r_g$ is the turbulence driving scale. We compare these two scales in the lower panel of Fig.~\ref{fig:cooling-in}. We observe that reconnection-driven acceleration is faster compared to stochastic acceleration, $t_{\rm r} \ll t_{\rm s} $, for $E_p<E_{\rm wall}$.
Below we constrain the spectra of protons accelerated by the two processes.

\paragraph{Proton acceleration by reconnection.} In this scenario, the corona is modeled as a collection of current sheets, that accelerate and confine pairs and protons. To study this case, we perform radiative PIC simulations of a current sheet, initialized in a Harris equilibrium, with the upstream plasma composed of pairs and ions \citep{chernoglazov+23}. The importance of IC cooling for leptons is set by ${\gamma_{\rm IC}}$, defined as the particle Lorentz factor for which cooling rate balances the acceleration rate from reconnecting electric fields. We set ${\gamma_{\rm IC}}/\sigma_{\pm}=10$, corresponding to a dynamically weak cooling expected in the AGN corona \footnote{See SM for additional discussion of simulation parameters \citep{tristanv2} and previous work on radiative reconnection \citep{hakobyan+19,zhang+21}}.
%More details are in the Supp. Material.

\begin{figure}
	\centering
	\includegraphics[width=0.48\textwidth,clip=false,trim= 0 0 0 0]{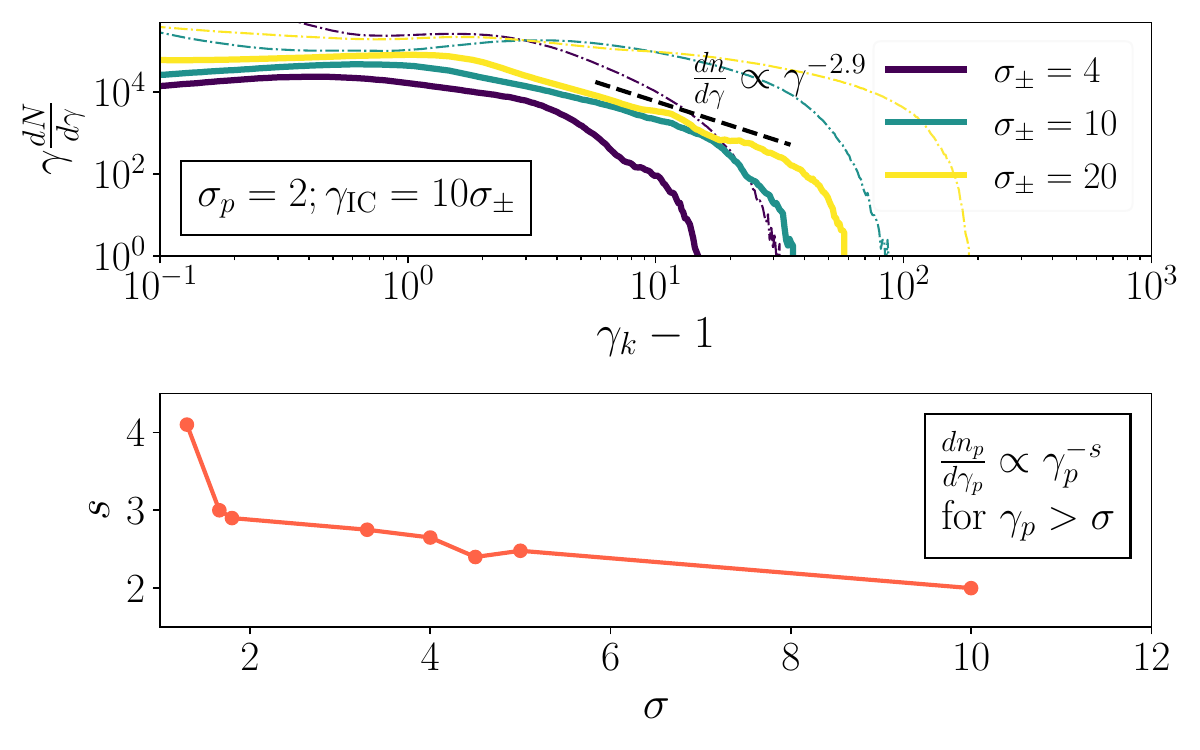}
	\caption{ Upper Panel: Spectral features of pairs (dot-dashed) and protons (solid) for cooled current sheets with $\gamma_{\rm IC} = 10\sigma_\pm$. Protons are accelerated to $\sigma_p$ with a flat spectrum and for $\gamma_p > \sigma$, the spectrum scales roughly as $\propto \gamma_p^{-s}$.
	Lower Panel: Empirical dependence of the spectral slope $s$ on $\sigma$ from PIC simulations.}
	\label{fig:spectra}
\end{figure}

In Fig.~\ref{fig:spectra}, we plot examples of proton spectra (solid lines) and pairs (dot-dashed), at 5 light-crossing times of the simulation box. We find that 
\emph{\textbf{i)}} $\gamma_{\rm IC}$ is the maximum attainable Lorentz factor for leptons.
%Pair acceleration is limited by radiation losses, such that 
%\emph{\textbf{iii)}} 
\emph{\textbf{ii)}} Proton acceleration is not affected by lepton cooling, and produces a spectrum containing most of the energy at $\gamma\sim \sigma \simeq \sigma_p$.
\emph{\textbf{iii)}} For $\gamma_p \gtrsim \sigma_p$, the proton spectrum is expressed as ${dn}/{d\gamma_p} \propto \gamma_p^{-s}$ (Upper Panel of Fig~\ref{fig:spectra}), where $s \in [2, 5]$ depending on $\sigma$ (Lower Panel of Fig.~\ref{fig:spectra}. See also \citep[e.g.][]{sironi+14,chernoglazov+23}).
%In general, $\sigma \sim 3$ is also expected if reconnection dominates particle acceleration.

\begin{figure}
	\centering
	\includegraphics[width=0.48\textwidth,clip=false,trim= 0 0 0 0]{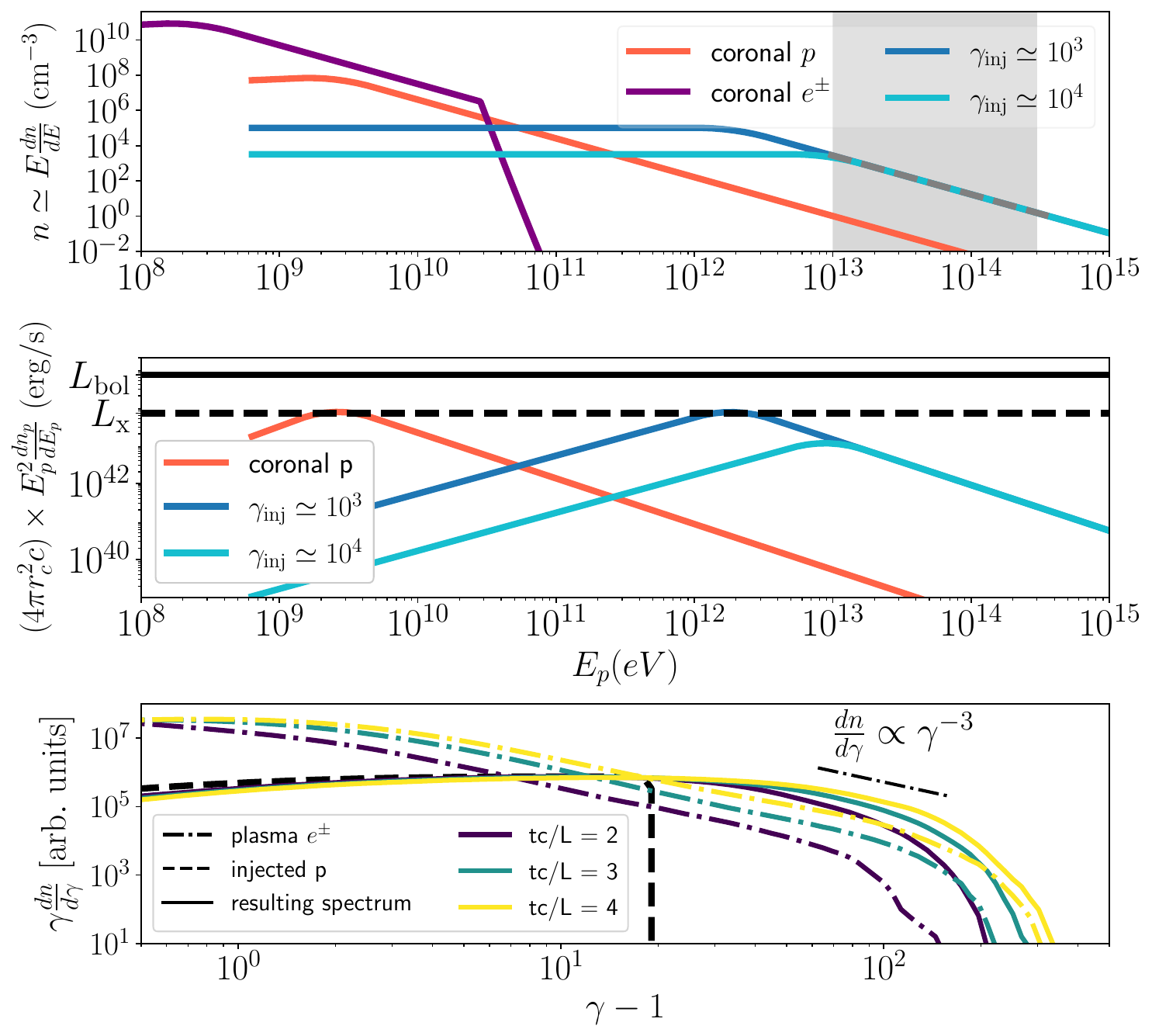}
	\caption{Upper Panel: Density of coronal protons and pairs, along with the proton spectral range set by $\gamma_{\rm inj}$ needed to explain the NGC-1068 signal \citep{IceCube-NGC1068}. Dashed line is the proton density associated with the same signal. Middle Panel: Luminosity requirements in the expected $\gamma_{\rm inj}$ range. Lower Panel: PIC simulation results from the injection of test protons with high $\gamma_{\rm inj}$ in magnetized turbulence with $\sigma =3$.}.
	\label{fig:np}
\end{figure}

\paragraph{Proton acceleration by turbulence.}
In this scenario, the corona is magnetically-dominated and turbulent with amplitude fluctuations, $\delta B \sim B_c$. 
We analyze 3D driven turbulence simulations \citep[setup based on work by][]{tenbarge+14,zhdankin+17,groselj+19,groselj+24}, until \rostom{$\sim 5$} light-crossing times of the simulation box.
In addition to initial thermal particles \footnote{Since we are interested in the re-acceleration of protons injected with large energies, such that their larmor radius is significantly larger compared to the plasma scales of the background plasma, we fix the mass ratio of the background species to be 1 in these simulations.}, we add a low-density population of ``pre-accelerated'' protons injected with a hard spectrum, ${dn}/{d\gamma_p}\sim \gamma^{-1}_p$ until $\gamma_p\sim \gamma_{\rm inj}$ \footnote{For simplicity, we considered an injection spectrum that truncates at $\gamma_{\rm inj}$. Dependence of the high-energy tail of the distribution function on the tail of the injected distribution will be the subject of future work}. A large population of reaccelerated particles is obtained, with a nonthermal slope, $s$, that hardens as $\sigma$ increases. 
%Following \citep{comisso+19}, we infer that $s \simeq 3$ is compatible with $\sigma \simeq 3$. 
We find that the energy-containing Lorentz factor of the evolved distribution of ``pre-accelerated'' particles remains at $\gamma_{\rm inj}$, along a tail with a slope $s \simeq 3$, similar to that of the background particles (lower panel in Fig.~\ref{fig:np}).

\paragraph{\bf Proton Acceleration Synthesis.} The most constraining property of the IceCube signal is its high luminosity. If we rely on particles accelerated from the coronal thermal pool, a significant proton luminosity at $\gamma_p \sim 10^3-10^4$ can only be achieved in the reconnection scenario for $\sigma_p \gtrsim 10$ ($s\approx 2$). This, however, yields a spectrum much harder than the observed slope of the neutrino spectrum.
We conclude that if starting from a thermal population, proton acceleration in the corona cannot explain the IceCube signal.

The scenario of turbulent confinement and re-acceleration of pre-accelerated protons can favorably compare with observations. The required slope at high energies, $s\sim 3$, is most consistent with re-acceleration in a coronal turbulent plasma of $\sigma\sim \sigma_p \gtrsim 1$. This corresponds to $\bar{n}_e/\bar{n}_p \approx 500$, favoring $p \gamma$ over $pp$ interactions (see Fig.~\ref{fig:cooling-in}). In Fig.~\ref{fig:np}, top panel, we plot the expected coronal proton densities, $\gamma_p (dn/d\gamma_p)$, (solid red lines) that we extend to NGC~1068's relevant energy range, and compare with the proton density $n \simeq E_p {dn_p}/{dE_p} \approx {L_p}/({4\pi \rc^2 c E_p})$ necessary to produce IceCube's signal (dashed grey from Eq.~\ref{eq:Lp}). The middle panel of Fig.~\ref{fig:np} constrains this scenario by requiring the total luminosity of injected protons $L_p$ to be below $\lx$, implying $\gamma_{\rm inj} \sim 10^3$. The lowest required proton luminosity, $L_p$, scales as $\sim \lx/10$, for the injection at $\gamma_{\rm inj} \sim 10^4$. The expected injected density is $\sim 10^{-5}-10^{-3}$ of $\bar{n}_p$. 

To that end, we propose that intermittently appearing current sheets in the vicinity of the BH, with magnetization $\sigma_{\rm p, bst}$ and the amount of dissipation of magnetic energy comparable to that happening in the corona, result in \emph{bursts} of a low-density population of protons impulsively accelerated up to $\gamma_{\rm inj} \sim \sigma_{\rm p, bst}$, and injected into magnetically-dominated and turbulent coronae, where protons are confined and re-accelerated. 
Such sheets could occur \emph{i)} at the disk/outflow boundary where relativistic asymmetric reconnection could be prominent \citep{mbarek+22}, i.e. one side of the flow containing a low proton density, 
\emph{ii)} at the BH's rotational equator during magnetic flux eruptions regulating the magnetic flux on the BH through episodically occurring magnetic reconnection \citep{ripperda+22},
or \emph{iii)} at the interface between magnetic loops of alternating polarity that are advected into the polar region, where the magnetization is relatively high \citep{chashkina+21,galeev+79,uzdensky+08,parfrey+15}
Such a configuration produces relatively weak BH outflows, that might be more compatible with radio observations of NGC 1068.
One of these scenarios or a combination thereof is plausible (See Appendix), as large scale general relativistic magnetohydrodynamic simulations of accretion flows around BHs show that \emph{a)} reconnection occurs at such boundaries \citep[e.g.][]{comisso+18,ripperda+20,sironi+21,mbarek+22}, and \emph{b)} magnetic flux eruptions occur in thin radiatively-efficient disks \citep[e.g.][]{scepi+22,liska+22}.

If the proton slope is steeper, $s \approx 3.5$, the non-relativistic, but still magnetically-dominated, high-amplitude turbulence can also serve as the re-acceleration mechanism \citep{comisso+22}. 
Turbulence becomes non-relativistic for $\sigma_p \lesssim 1$, corresponding to  $\bar{n}_e/\bar{n}_p \lesssim 150$. This scenario requires the re-acceleration length, $\sim 3 r_c (0.1/\sigma)$, to be shorter than the proton cooling time, $\lambda_{\nu}$. At low enough $\sigma_p$ such that $\bar{n}_p \approx \bar{n}_e$, the cooling is dominated by $pp$, such that $\lambda_{\nu}\sim 10 r_c$. If $ct_{\rm s}\lesssim \lambda_{\nu}$ then $\sigma \gtrsim 0.03$. Additional constraints can be provided by Bethe-Heitler cooling (See SM).

\paragraph{\bf Conclusions.} We present basic steps connecting first-principles plasma simulations of proton acceleration in AGN coronae to the observed interplay of X-rays, $\gamma$-rays, and neutrinos. Our most robust conclusions are that \emph{\textbf{i})} radiation fields in the corona lead to efficient absorption of hadronic $\gamma$-rays within 100 $r_g$ of the BH; \emph{\textbf{ii})} protons accelerated from the coronal thermal pool cannot account for NGC~1068's neutrinos; \emph{\textbf{ii})} the observed neutrinos stem from $p\gamma$ interactions; \emph{\textbf{iv})} explaining the neutrino signal requires injection of protons pre-accelerated to $\gamma\sim 10^3-10^4$, into the turbulent magnetically-dominated corona, where they are confined and reaccelerated. 

\section*{Acknowledgements}

We would like to thank Irene Tamborra for providing important insights on neutrino and $\gamma$-ray production, Hayk Hakobyan for valuable help in setting up PIC simulations, and Bart Ripperda, Daniel ~Gro\v selj, Lorenzo Sironi, Kohta Murase and Eliot Quataert for useful conversations. This work was supported by a grant from the Simons Foundation (MP-SCMPS-00001470) to AP and AL, and facilitated by Multimessenger Plasma Physics Center (MPPC), NSF Grant No. PHY-2206610. Computing resources were provided by the Division of Information Technology at the University of Maryland {\texttt{Zaratan} cluster} \footnote{\url{http://hpcc.umd.edu}}. This research is part of the Frontera computing project at the Texas Advanced Computing Center (LRAC-AST21006). Frontera is made possible by NSF award OAC-1818253.

\appendix
\section*{Appendix on Radiation Fields, Confinement, and Potential injection Scenarios.}

\subsection*{Radiation Fields}

\paragraph{Optical/UV (OUV):}
A tight relation between the OUV and X-ray radiation is observed in AGNs, and is set by the spectral index $\alpha_{\rm ox} = -{\log{(L_{\rm 2keV}/L_{\rm 2500 \text{\AA}})}}/{2.605}$ \citep[e.g.,][]{tananbaum+79,zamorani+81,silverman+05,steffen+06,just+07}.
The integrated intrinsic X-ray luminosity of NGC~1068 in the 2-10keV band is $L_{\rm x} = 7^{+7}_{-4} \times 10^{43}\ergs$ \citep{marinucci+16}. 
We can then estimate the luminosity $L_{\rm 2500 \text{\AA}}$ \citep{lusso+10,lusso+12}:
\begin{equation}\label{eq:Lxuv}
	\log{L_{\rm 2keV}} = (0.760 \pm 0.022) \log{L_{\rm 2500 \text{\AA}}} + (3.508 \pm 0.641)
\end{equation}

We estimate the OUV energy density based on coronal X-rays as $U_{\rm ouv} = {L_{\rm 2500 \text{\AA}}}/{(4 \pi c r_{\rm ouv}^2)}$, where $r_{\rm ouv}$ defines the radius within which half of the observed light is contained (the half-light radius). From quasar microlensing observations along with the AGN disk size estimates, $r_{\rm ouv} \sim 100 r_g$ over a wide range of BH masses $\gtrsim 10^7 M_{\odot}$ \citep[e.g.,][]{edelson+15}. Seyfert galaxies have quasar-like nuclei, but with a clearly detectable host galaxy, thus, we rely on quasar continuum models \citep{zheng+87} to obtain ${dn_{\rm ouv}}/{d \epsilon} \approx U_{\rm ouv} \epsilon^{-1}$ for [1050, 2500]~\AA. 

\paragraph{X-rays:}
X-rays are the dominant targets for the $p\gamma$ process as the lowest energy photons, $\epsilon_{\rm min}$, that contribute to producing $E_{\nu}\sim$10 TeV neutrinos have energies $\epsilon_{\rm min} = ({\alpha m_p c^2 \bar{\epsilon}_{\rm th}})/E_{\nu} \simeq 1$~keV, where $m_p$ is the proton mass, and  $\bar{\epsilon}_{\rm th} \approx 0.15$ GeV is the $p\gamma$ energy threshold in the proton frame. For NGC~1068, we deduce a spectral energy density $U_{\rm x} \epsilon_0^{-2} \text{if } \epsilon < 20 \text{ keV}$ and $U_{\rm x} \epsilon^{-2} \text{if }  20 \leq \epsilon \leq 200 \text{ keV}$ \citep[Fig.~2 in][]{bauer+15},
where $U_{\rm x} = {L_{\rm x}}/{(4 \pi \rc^2 c)}$ is the coronal X-ray energy density, $\rc \simeq 10 r_g$ is the corona size \citep[e.g.][]{dai+10,fabian+15}, and $\epsilon_0 \simeq 7$keV \footnote{$\epsilon_0$ corresponds to the broad iron K$\alpha$ emission for NGC~1068---where most of the power lies---originating from the central regions of the black hole accretion disk \citep[e.g.][]{reynolds+99}.}. We note that the X-ray spectral shape of these obscured sources is uncertain and the spectrum of type I Seyferts might be different.

\begin{figure}
	\centering
	\includegraphics[width=0.45\textwidth,clip=false,trim= 0 0 0 0]{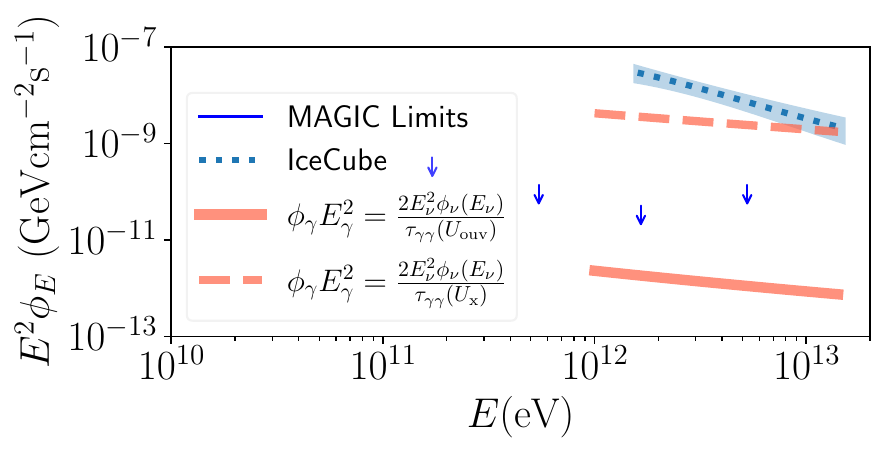}
	\caption{$\gamma$-ray spectrum suppression based on X-rays and OUV (Eq.~\ref{eq:gamma}), along with IceCube\citep{IceCube-NGC1068}, and MAGIC \citep{MAGIC19} limits in the NGC~1068 signal energy range. X-rays cannot account for the $\gamma$-ray suppression above $\sim 10$ GeV, but the 2500\AA~emission provides enough suppression.}
 \label{fig:gamma}
\end{figure}

\subsection*{$\gamma$-ray Interactions}
High-energy photons with energies $\epsilon_{\gamma}$ interact most efficiently through Breit-Wheeler \citep{breit+34} with target photons of energy $\epsilon_{\rm t} \sim m_e^2c^4/\epsilon_{\gamma}$, which gives $\epsilon_{\rm t} > 0.1 $~eV for $\sim$TeV $\gamma$-rays. We calculate the center of momentum energy of target photons with energy $\epsilon_t$ and high-energy photons $\epsilon_\gamma$, such that $ \mathcal{S} = {\epsilon_t \epsilon_\gamma }(1 - \cos{\theta})/({2 m_e^2 c^4})$, where $\theta$ is the angle between the momenta of photons in the laboratory frame. Assuming isotropic target photon field, one can estimate $ \mathcal{S} \approx {\epsilon_t \epsilon_\gamma }/({2 m_e^2 c^4})$. Finally, the $\gamma\gamma$ cross-section, $\Sigma_{\gamma \gamma}$, can be expressed as:
\begin{equation}\label{eq:sig-gg}
	\Sigma_{\gamma \gamma} = \frac{\pi}{2} r_e^2 (1 - \phi^2) \big[ -2 \phi (2 - \phi^2) + (3 - \phi^4) \ln{\frac{1+\phi}{1-\phi}} \big]
\end{equation}
where $ \phi = \sqrt{1 - 1/\mathcal{S}}$. We can then calculate the $\gamma \gamma$ optical depth for a coronal region of size $r_c \sim 10r_g$. 

The optical depth for $\gamma \gamma$ is, $\tau_{\gamma \gamma} (\epsilon_{\gamma}) \simeq \int_{\epsilon_{\rm t,min}}^{\epsilon_{\rm t,max}} \rc \Sigma_{\gamma \gamma}(\epsilon_{\rm t}, \epsilon_\gamma)  ({dn_{\rm t}}/{d\epsilon_{\rm t}})  d\epsilon_{\rm t} $, where $\Sigma_{\gamma \gamma}(\epsilon_{\rm t}, \epsilon_\gamma)$ is the cross section for $\gamma \gamma$, and $\epsilon_{\rm t}$ is the  target photon energy. 
The resulting optical depth is shown in Fig.~2 in the SM.

Since $\gamma$-rays are produced and absorbed throughout the corona, the $\gamma$-ray luminosity in the IceCube energy range is,
\begin{equation}\label{eq:gamma}
	\epsilon_\gamma^2 \phi_\gamma(\epsilon_\gamma) \sim \frac{2 E_\nu^2 \phi_\nu(E_\nu)}{\tau_{\gamma \gamma}} (1 - e^{-\tau_{\gamma \gamma}}).
\end{equation}
We finally obtain Fig.~\ref{fig:gamma}, which shows the impact of $\gamma \gamma$ interactions on the $\gamma$-ray flux.
Interactions with both OUV and X-rays are considerable, for $\epsilon_\gamma \lesssim 10{\rm TeV}$ (See Fig.~\ref{fig:gamma}). However, OUV photons are needed to curb the $\gamma$-ray flux at the TeV level.

\subsection*{Photomeson Cooling}

To calculate $\lambda_{p \gamma}=c/t^{-1}_{p \gamma}$ for protons with Lorentz factor $\gamma_p$, we employ \citep{stecker68}, 
\begin{equation}\label{eq:s0}
	t^{-1}_{p \gamma}(\gamma_p) \approx \frac{\xi c}{2 \gamma_p^2} \int^{\infty}_{\bar{\epsilon}_{\rm th}/2\gamma_p} d\epsilon \frac{dn_{\rm x}}{d\epsilon} \epsilon^{-2}  \int^{2 \gamma_p \epsilon}_{\bar{\epsilon}_{\rm th}}  \epsilon ' \Sigma_{\rm eff}(\epsilon ') d\epsilon ',
\end{equation}
where $\epsilon ' \approx 2 \gamma_p \epsilon$ is the photon energy in the proton rest frame, $\epsilon$ is its energy in the BH frame, $\xi$ is the inelasticity, and $\Sigma$ is the $p\gamma$ cross section.

\paragraph{X-rays:} Based on NGC~1068's X-ray spectral features \citep{bauer+15}, an effective cross section $\Sigma_{\rm eff} \simeq 150 \mu$b \citep{PDG18}, and an interaction threshold $E_p =m_p c^2 \bar{\epsilon}_{\rm th}/(2 \epsilon_0) \gtrsim 10^{13}$eV, Eq.~\ref{eq:s0} yields
\begin{equation}\label{eq:photomes}
	\frac{t^{-1}_{p \gamma}(\gamma_p)}{\frac{4}{3} \xi c \Sigma_{\rm eff}}  = 
	\begin{cases}
		 \frac{ U_{\rm x}}{2 \epsilon_0} (3 + \frac{ \bar{\epsilon}_{\rm th}^2 }{ 4 \epsilon_0^2 \gamma_p^2 } - 3 \frac{\bar{\epsilon}_{\rm th}}{2 \epsilon_0 \gamma_p}),& \text{if } \gamma_p > \bar{\epsilon}_{\rm th}/(2 \epsilon_0)\\
		 U_{\rm x} \frac{\gamma_p}{\bar{\epsilon}_{\rm th}},& \text{if } \gamma_p < \bar{\epsilon}_{\rm th}/(2 \epsilon_0)
	\end{cases}	
\end{equation}

\paragraph{OUV:} We calculate $t^{-1}_{p \gamma}$ for the OUV component based on Eq.~\ref{eq:s0}, where ${dn_{\rm ouv}/}{d \epsilon} \approx U_{\rm ouv} \epsilon^{-1}$ in the 1050-2500~\AA~range. In the vicinity of the corona, the photon energy density is $U_{\rm ouv} = {L_{\rm 2500 \text{\AA}}}/({4 \pi c r_{\rm ouv}^2})$, and we get for $\gamma_p > \bar{\epsilon}_{\rm th}/(2 \epsilon_c)$,
\begin{equation}\label{eq:OUV}
	t^{-1}_{p \gamma}(\gamma_p)  = \xi c \Sigma_{\rm eff} U_{\rm ouv} \left(\ln{\frac{\gamma_p}{\gamma_c}} + \frac{\gamma_c^2}{2 \gamma_p^2} - \frac{1}{2}\right),
\end{equation}
where $\epsilon_c\approx 12$eV (1050\AA) is the photon energy.

\subsection*{Confinement of High-Energy Protons}
\paragraph{Confinement in the corona}
As discussed above, protons are confined in the corona before interacting with radiation fields to produce neutrinos. 
In Fig.~4 of the SM, we show the ratio of $t_{p\gamma}/t_{\rm conf}$ and $t_{pp}/t_{\rm conf}$, where  $t_{\rm conf}$ is the proton confinement time defined as $t_{\rm conf} = \text{min}(t_{\rm esc}(B_c), \rc/v_{\rm io}) $, with $v_{\rm io}$ the speed of a potential coronal inflow/outflow.

The first conclusion we draw is that \emph{i)} protons are confined for long enough to efficiently interact through $p\gamma$ for the considered radiation fields, and through $pp$ if $\bar{n}_e/\bar{n}_p \lesssim 100 $. Note that since particles are most efficiently confined in the diffusive regime as set by Eq.~\ref{eq:tesc} in the main text, $\bar{n}_e/\bar{n}_p \simeq 100 $ is the maximum allowed ratio for $pp$ interactions to occur in the corona, irrespective of the radiation fields.
The second conclusion is that \emph{i)} if either an inflow, (part of the corona is accreting into the BH), or an outflow (e.g. a wind launched from the corona) with $v_{\rm io} \gtrsim 10^{-2}c$ is present (See Eq.~\ref{eq:tesc}), neutrino production efficiency can be suppressed as the cooling time becomes longer than the confinement time, $t_{\nu}/t_{\rm conf} \gtrsim 1/3$.

\paragraph{Confinement in a 3D magnetically-dominated turbulence box.}
To test the confinement process, we track injected protons and check whether their mean free path, $\lambda_{\rm bst}$, is consistent with the expected random walk confinement estimate, $\lambda_{\rm esc}\simeq r_g (\rl/r_g)^{1/3}$ (Eq.~\ref{eq:tesc}). 
In coronae, we generally expect $\lambda_{\rm esc}/r_c \simeq 0.1 (\rl/r_g)^{1/3}$. In the energy range of interest and for $\sim 10^3$G field, $ \langle \lambda_{\rm esc}/r_c \rangle  \sim 5 \times 10^{-3}$. For our simulation setup, we expect $ \langle \lambda_{\rm c}/L \rangle  \simeq (l/L) (\rl^c/l)^{1/3}$, where $\rl^c$ is the energy-dependent proton Larmor radius in our simulations. For our simulation parameters, we obtain $ \langle \lambda_{\rm c}/L \rangle \sim 10^{-2}$.

We finally calculate the mean free path of test protons, $\lambda_{\rm bst}$, in our simulations if they follow a random walk. We calculate the displacement of particles $d$ and the total distance traveled $D$, such that $\lambda_{\rm bst} \simeq d^2/D$. We show the obtained results in Fig~5 in the SM, where we find $ \lambda_{\rm bst}/L \in [10^{-3},10^{-2}] $.
Therefore, we find that $ \lambda_{\rm bst}/L \sim \lambda_{\rm c}/L \sim  \lambda_{\rm esc}/r_c$, and show that the mean free path estimates are consistent with our PIC simulations. A more detailed analysis of energetic proton confinement in magnetically-dominated turbulence in PIC simulations will be the subject of a forthcoming study.

\subsection*{Potential Injection Scenarios}
This Letter offers strong hints that---provided IceCube's signal originates from the corona of NGC 1068---an injection of nonthermal particle populations, not accelerated in the coronal plasma and responsible for high-energy neutrinos, is occurring. This is an important point in the plasma and particle acceleration communities, as it shapes our understanding of the structure of the corona itself and, potentially, the accretion flow and magnetic field structure in the vicinity of the BH.

The exact mechanism injecting protons at $\gamma_p \simeq 10^3-10^4$ is unclear. However, we present a few ideas that are worth expanding on.
For instance, we put forward that asymmetric reconnection that self-consistently occurs at the boundary of the turbulent accreting plasma could control the injection up to $\gamma_p \sim \sigma_p$. Further acceleration would happen because of stochastic acceleration in turbulence. An analogy could be drawn here from turbulent acceleration in a magnetized plasma, where injection of accelerated particles occurs through reconnection, followed by re-acceleration through turbulence in the same magnetically-dominated region \citep{comisso+18,comisso+19}. In our scenario, the initial acceleration occurs through reconnection in a significantly more magnetized plasma, and subsequent stochastic acceleration in the corona. Ultimately, the slope of the power-law tail at the highest energies is dictated by turbulence, instead of reconnection, as shown in our simulations. Self-consistent PIC simulations are needed to investigate the interplay of asymmetric reconnection, occurring at the boundary of media with different magnetizations, and turbulence.

Alternative scenarios we put forward include magnetic flux eruptions in arrested accretion \citep{ripperda+22}, and acceleration at the interface between magnetic loops of alternating polarity, which can form in the corona \citep[e.g.][]{galeev+79,uzdensky+08,parfrey+15}. Both of these alternatives include initial acceleration through reconnection followed by re-acceleration in the magnetized turbulence.
In both instances, reconnection can be occurring in the regime of a low guide field, and, with laminar initial conditions, producing a hard power-law tail with $s \sim 2$ \citep[e.g.][]{sironi+14,chernoglazov+23}. However, an injected population with $s \sim 2$ could experience a steepening to $s \sim 3$ because of turbulent re-acceleration if $\rl< l$, where $l$ is the energy-carrying scale in the turbulent medium. 
We were not able to fully test this hypothesis in our PIC simulations because of a limited separation of scales; however, we expect injected particles to be stochastically reaccelerated in turbulent media. 
We will investigate stochastic reacceleration more thoroughly in a forthcoming publication. An additional possibility consists in current sheets possessing a finite guide field, which could steepen the spectrum \citep{fiorillo+23b}.

\bibliography{Total}

%%%%%%%%%%%%%%%%%%%%%%%%%%%%%%%
%%%%%%%%%%%%%%%%%%%%%%%%%%%%%%%
%%%%%%%%%%%%%%%%%%%%%%%%%%%%%%%
%%%%%%%%%%%%%%%%%%%%%%%%%%%%%%%
%%%%%%%%%%%%%%%%%%%%%%%%%%%%%%%
%%%%%%%%%%%%%%%%%%%%%%%%%%%%%%%
\section*{Supplementary Material (SM)} 

\subsection{Bethe-Heitler Interactions}
We also consider the effects of Bethe-Heitler interactions ($p+\gamma \rightarrow p e^+ e^-$), which can be important for modeling the cooling of coronal protons \citep{murase+20}. The Bethe-Heitler (bth) cooling time can be expressed as,
\begin{equation}\label{eq:BH}
	t^{-1}_{\rm bth}(\gamma_p) \approx \frac{c}{2 \gamma_p^2} \int^{\infty}_{\bar{\epsilon}_{\rm bth}/2\gamma_p} d\epsilon \frac{dn_{\rm x}}{d\epsilon} \epsilon^{-2}  \int^{2 \gamma_p \epsilon}_{\bar{\epsilon}_{\rm bth}}  \epsilon ' \Sigma(\epsilon ') d\epsilon ',
\end{equation}
where $\bar{\epsilon}_{\rm bth} = 2$MeV is the threshold for Bethe-Heitler interactions in the proton frame.

\begin{figure}
	\centering
	\includegraphics[width=0.48\textwidth,clip=false,trim= 0 0 0 0]{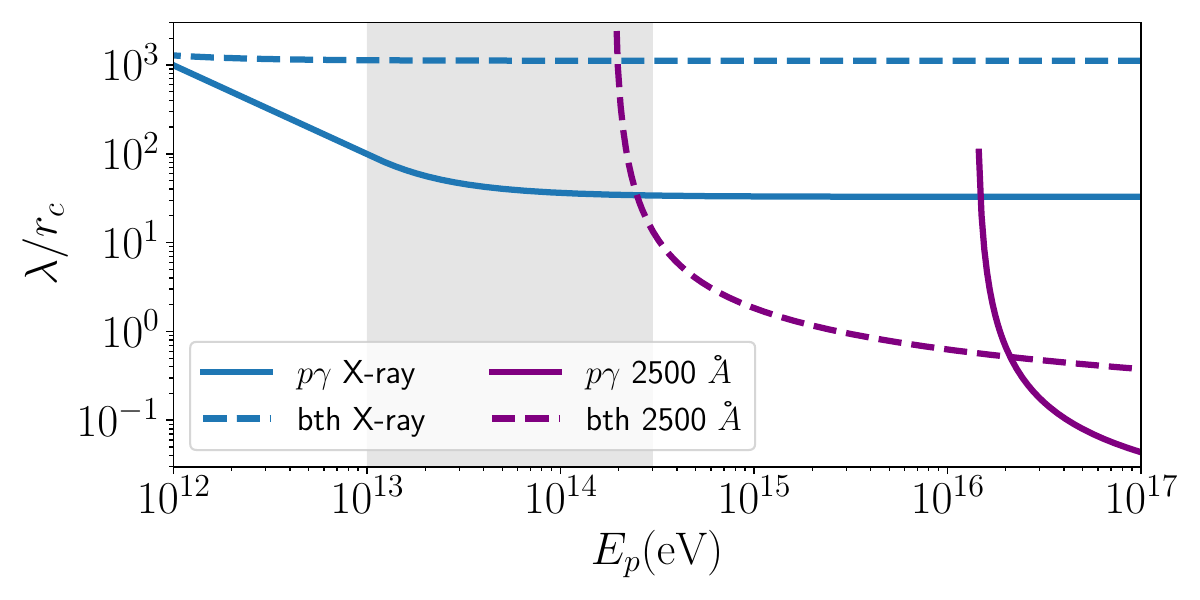}
	\caption{Proton mean free path to Bethe-Heitler interactions based on the photon fields considered in this letter. In the energy range relevant for NGC~1068 neutrinos, $p\gamma$ interactions are more efficient than bth interactions.}
	\label{fig:BH}
\end{figure}

We plot the corresponding mean free path in Fig~\ref{fig:BH}. For the photon fields considered in this study, $p\gamma$ effects are more prominent in the energy range of interest to NGC~1068. 
The effects shown in Fig~\ref{fig:BH} are highly dependent on the assumed photon spectral shape, which is highly uncertain and might be different for different sources.
We note that Bethe-Heitler interactions could have stronger effects on the available proton population, if we take into account far UV photons (obscured by the interstellar medium) potentially produced by the accretion disk in the energy range $\in [20, 200]$~eV.

\subsection{Pair Cooling}
\paragraph{Inverse Compton cooling.} Electrons and positrons in the black hole corona experience efficient inverse Compton cooling. It is convenient to parameterize it through the radiation-reaction limit, $\gamma_{\rm IC}$, which corresponds to the Lorentz factor for which the energy gain due to acceleration by the reconnecting electric field is balanced by cooling through inverse Compton:
\begin{equation}\label{eq:gic1}
	\gamma_{\rm IC}^2 \simeq \frac{e E}{\Sigma_T U_\gamma} \sim \frac{e \beta_{\rm r} \rc B_c}{\ell m_e c^2} = \frac{\beta_r \rc }{\ell \rl}
\end{equation}
where $U_\gamma$ is the energy density of the photon field,  $\ell  = {(\Sigma_T U_\gamma \rc)}/({m_e c^2})$ is the radiative compactness parameter, and $\rl=m_e c^2/(eB_c)$ is the nominal Larmor radius. It can be expressed as $\rl = d_e/\sqrt{\sigma_\pm}$, where $d_e$ is the plasma skin depth. Thus,
\begin{equation}\label{eq:gic2}
	\gamma_{\rm IC} \simeq \sqrt{\frac{\beta_r\sigma_\pm}{\ell} \left(\frac{r_c}{d_e}\right) }
\end{equation}
For the parameters of NGC~1068, we obtain $\gamma_{\rm IC} \sim 10^5 \gg \sigma_{\pm}$. As a result, inverse Compton cooling does not impact quick reconnection-driven acceleration of leptons until $\gamma\sim \gamma_{\rm IC}$. However, the cooling time is still sufficiently smaller than the light-crossing time of the corona, $t^{e}_{\rm IC}/(r_c/c)=(3/4) (\ell \gamma)^{-1}\ll 1$, which should severely limit lepton energization through stochastic acceleration.

\paragraph{Synchrotron cooling.} Similar estimates can be obtained for the radiation-reaction limit to synchrotron cooling. However, one should take into account that self-absorption blocks synchrotron cooling until $\gamma\approx 100\gtrsim \sigma_{\pm}$ \citep{beloborodov17}.

\subsection{Synchrotron Cooling}

Since the magnetic fields in black hole coronae can be quite strong, we discuss the effects of synchrotron cooling.

\paragraph{Pion Cooling.}
In $p \gamma$ and $pp$ interactions, charged pions, $\pi^{\pm}$, decay to produce neutrinos,
\begin{equation}
	\pi^{\text{+}} \rightarrow e^{\text{+}}+\nu_e+\nu_\mu+ \bar{\nu}_\mu,
\end{equation} 
with a decay time $\tau_\pi = 2.6033 \times 10^{-8}$s in the rest frame.

The synchrotron cooling time for pions is $t_{\pi, s}(E_\pi) = {3 m_\pi^4 c^7}/({2 e^4 B^2_c E_\pi})$. For neutrinos to be produced, $\gamma_\pi \tau_\pi < t_{\pi, s}$ must be satisfied \citep[e.g.,][]{guarini+23}. We can see that for $B_c\sim 10^4$G, pions decay before cooling through synchrotron in the energy range of interest.  It is worth noting that if coronae of X-ray binaries, where magnetic fields are much stronger, $B_c \sim 10^7$G, were neutrino emitters, these effects could be crucial in determining the maximum energies of emitted neutrinos. Potential neutrino emission from X-ray binaries will be the topic of a forthcoming study.

\paragraph{Proton radiation-reaction limit.}

The cooling time for proton synchrotron is $t^p_{\rm syn} = {6 \pi m_p^4 c^3}/({\Sigma_T m_e^2 E_p B^2_c})$, which results in the following synchrotron radiation-reaction limit,
\begin{equation}
	\gamma^p_{\rm burn} = \sqrt{\frac{6 \pi m_p^2 \beta_r e}{\Sigma_T m_e^2 B_c}} \sim 2.15 \times 10^{9} \sqrt{{\beta_r}\left(\frac{10^4{\rm G}}{B_c}\right)}.
\end{equation}
Therefore, the radiation reaction limit is not significant for limiting the proton energy in black hole coronae.

{\paragraph{Proton Cooling time.} The cooling time for proton synchrotron can be expressed as
\begin{equation}\label{eq:tsyn}
\frac{t^p_{\rm syn}}{r_c/c} = \frac{3}{4}\left(\frac{m_p}{m_e}\right)^3 \frac{1}{\ell_B \gamma_p}\approx 500 \left(\frac{100}{\ell_B}\right) \left(\frac{10^5}{\gamma_p}\right),
\end{equation}
where $\ell_B = {\Sigma_T U_B \rc}/({m_e c^2})\sim 10-100$ is the magnetic compactness parameter. One can see that for $\gamma_p\sim 10^5$, the synchrotron cooling time is substantially longer than the stochastic acceleration time, $0.3 r_c/\sigma$ (see main text), and longer than the cooling time to $p\gamma$ via interaction with coronal X-rays. However, at energies just below the OUV wall, $\gamma_p \gtrsim 10^6-10^7$, synchrotron cooling becomes faster than the cooling to $p\gamma$.
}

\subsection{Optical Depth of $\gamma$-rays}
\begin{figure}
	\centering
	\includegraphics[width=0.48\textwidth,clip=false,trim= 0 0 0 0]{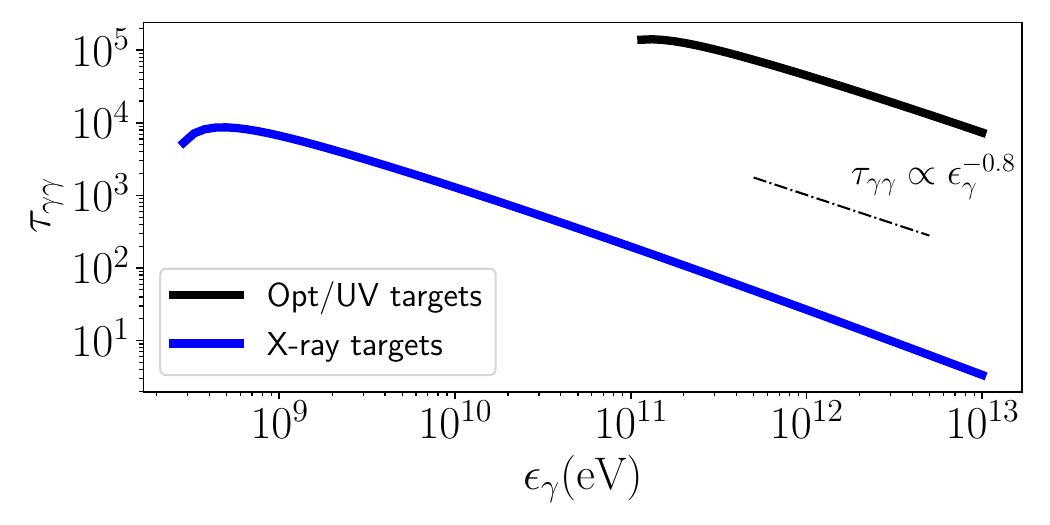}
	\caption{Optical depth of $\gamma\gamma$ interactions for pionic $\gamma$-rays depending on the soft photon target. }
	\label{fig:tau-gg}
\end{figure}

The optical depth of Breit-Wheeler interactions for $\gamma$-rays is shown in Fig~\ref{fig:tau-gg}, depending on the target photon field.

\subsection{3D PIC Simulations of Radiative Magnetic Reconnection}
There has been a substantial set of studies on radiative cooling in reconnecting  current sheets \citep[e.g.][]{beloborodov17}. Both synchrotron and inverse Compton losses have been shown to affect both the non-thermal particle acceleration \citep[e.g.,][]{hakobyan+19}, and the reconnection dynamics (for example, strong cooling in the absence of guide field leads to intensive compression of plasmoids because of the decrease in plasma temperature). In 3D current sheets, particle acceleration is not explored as thoroughly, especially in the radiative regime. In addition, not as much attention has been given to acceleration of ions to high energies, $\gamma \gtrsim \sigma$. 3D reconnection has been shown to be significantly more efficient than its 2D counterpart, as particles can escape the flux tubes and get directly accelerated by the larger-scale reconnecting field \citep{zhang+21}. Here, we perform 3D simulations that include inverse Compton cooling on pairs in 3D reconnection \citep{chernoglazov+23}, and a population of uncooled heavier ions, to investigate the properties of particle reconnection during magnetic reconnection in the corona. We consider dynamically weak inverse Compton cooling, $\gamma_{\rm IC} \gg \sigma_{\pm}$, representative of AGN corona.

\begin{figure}
	\centering
	\includegraphics[width=0.48\textwidth,clip=true,trim= 0 0 0 0]{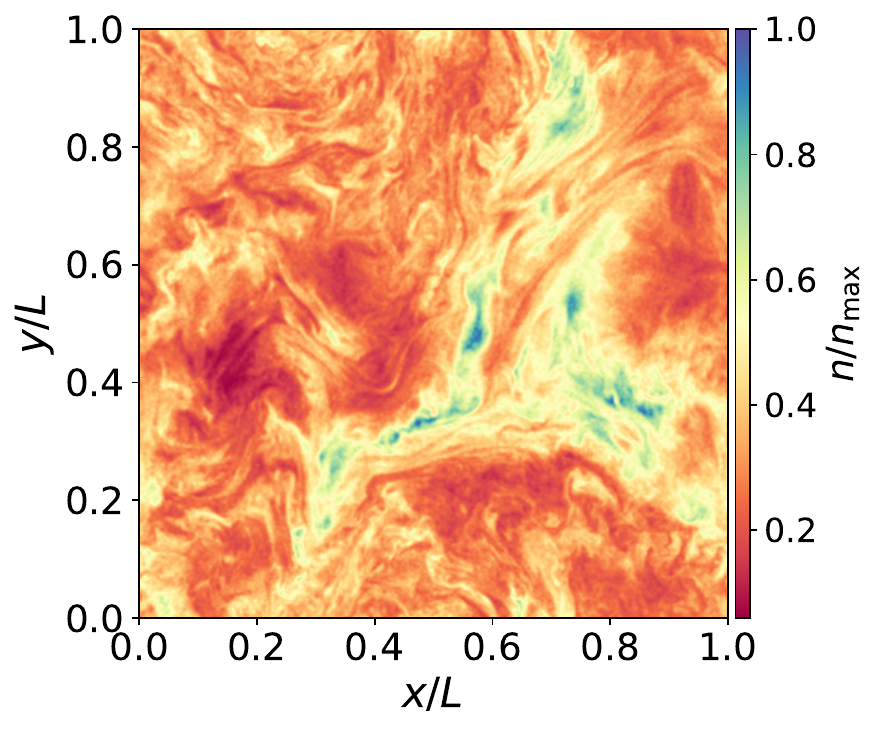}
	\caption{2D slice of Normalized number density map $n/n_{\rm max}$ in the 3D magnetically dominated turbulence simulation with $\sigma = 3$ and $\langle \delta B^2 \rangle / B_0^2 = 1$ after $\sim 3$ light-crossing time. $l = L/8$ is the energy-carrying scale which should scale with $\sim r_g$ in the corona.}
	\label{fig:density}
\end{figure}

\paragraph{Simulation Setup.}
We perform 3D  particle-in-cell (PIC) simulations of relativistic reconnection with both ions and electron-positron pairs using TRISTAN-MP v2 code \citep{tristanv2}. Our simulation setup closely follows \citet{chernoglazov+23}. A current sheet is initialized in a Harris equilibrium with a magnetic field strength profile $B = B_0 \hat{x} \tanh({z/\delta})$, where $\delta=7 d_e$ is the half-width of the current sheet and $x$ and $z$ are the coordinates along and across the magnetic field, respectively. We use the ion-to-electron mass ratio of $m_i/m_e = 25$. For relativistic reconnection, $\sigma_p \gtrsim 1$, ions are accelerated to high energies, and the exact value of the ion to electron mass ratio should not have much of an impact. The pairs upstream of the current sheet are initialized with a small temperature, $ k_BT/mc^2 = 10^{-2} $. We vary the electron-to-ion number density ratio, $n_e/n_i$, to study the effect of different pair loading. The simulations' length in the $x$, $y$, and $z$-directions is $1000~d_e$, where $d_e = c/\sqrt{4 \pi n_e e^2/m_e}$ is the electron skin depth. The acceleration to large energies happens through a reconnection-driven electric field \citep{chernoglazov+23}, which is in the $y$-direction. We use outflow boundary conditions for electromagnetic fields and particles in $x$ direction, periodic boundary conditions in the $y$ direction and injection boundary conditions in the $z$ direction. Each simulation uses one positron+electron+ion particle per cell and 1 cell per $d_e$. A convergence study was performed and showed that adding more particles per cell does not change the results. The Courant number was set to $c\Delta t/\Delta x = 0.25$.

\begin{figure}
	\centering
	\includegraphics[width=0.48\textwidth,clip=false,trim= 0 0 0 0]{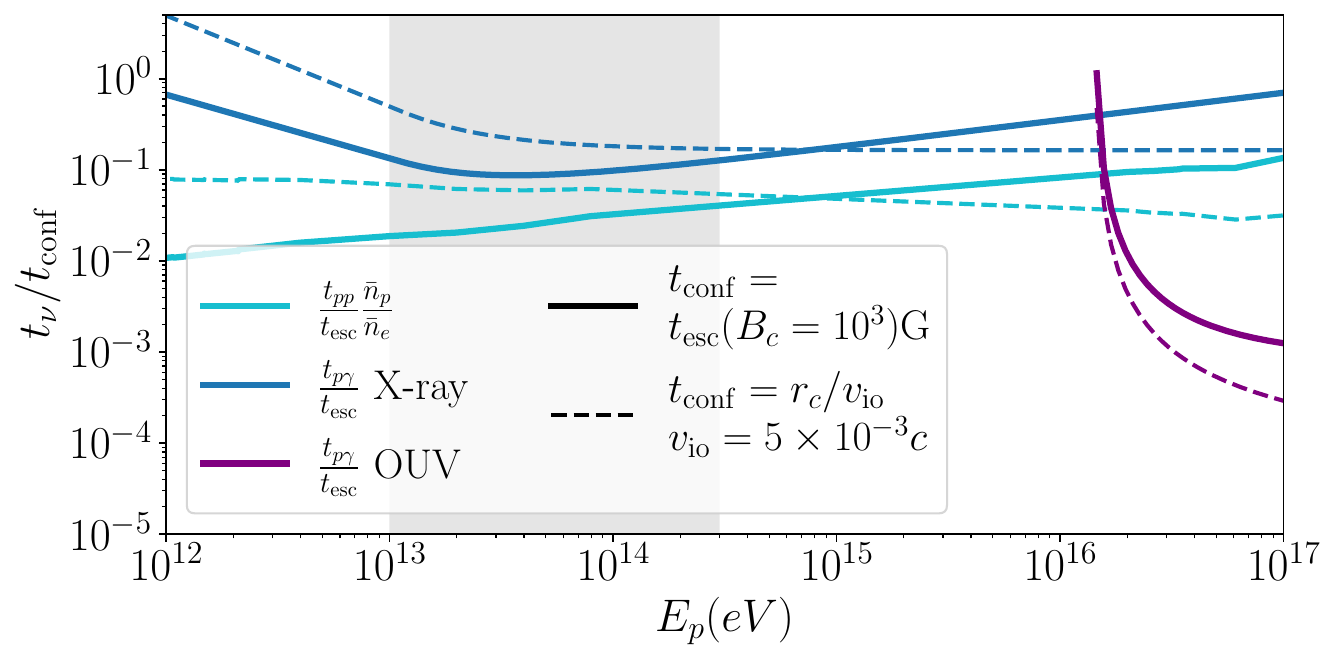}
	\caption{Ratio of $p\gamma$ and $pp$ cooling time, $t_\nu$, to proton confinement time, $t_{\rm conf} = \text{min}(t_{\rm esc}(B_c), \rc / v_{\rm io})$. $t_{\rm esc}$ is given by Eq.~\ref{eq:tesc} due to diffusion, and $v_{\rm io}$ is the inflow/outflow speed of the corona. This is plotted in the proton energy range responsible for NGC~1068 neutrinos (shaded region).}
	\label{fig:tnu-tesc}
\end{figure}

\begin{figure}
	\centering
	\includegraphics[width=0.48\textwidth,clip=true,trim= 0 0 0 0]{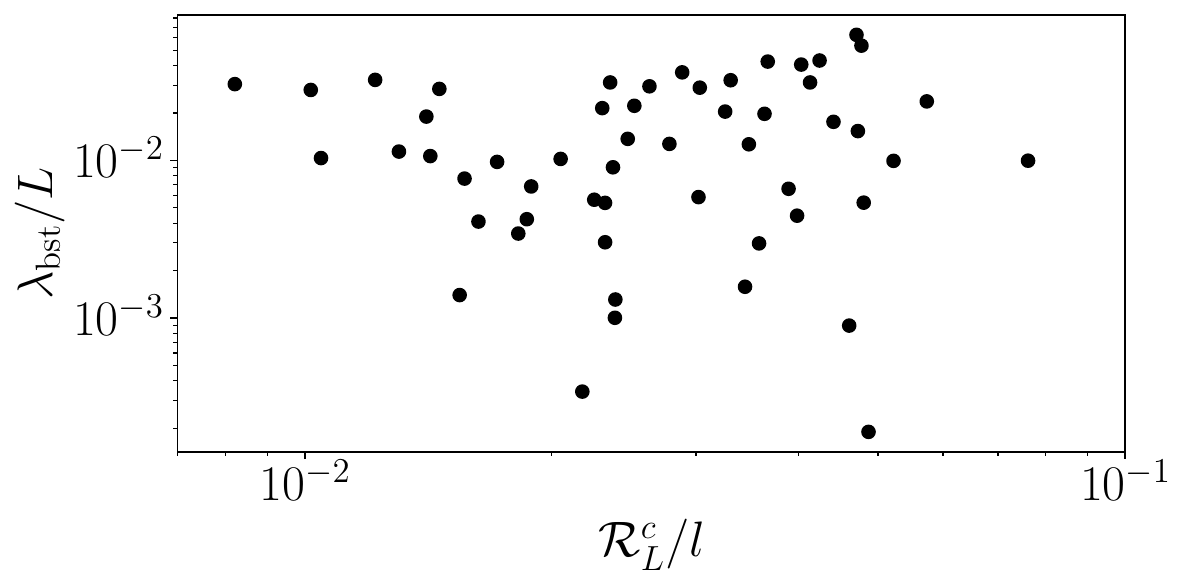}
	\caption{Mean free path $\lambda_{\rm bst}$ of injected particles from the proton burst in the turbulent box from Fig~\ref{fig:density} as a function of the normalized Lamor radius $\rl^c/l$, where $\rl^c = \gamma mc^2/B$. }
	\label{fig:confinement}
\end{figure}

\subsection{3D PIC Simulations of Magnetically-Dominated Turbulence}

\paragraph{Simulation Setup}
Our 3D PIC setup, i.e., simulation tracks all three components of electromagnetic field components,  particle velocities, and gradients, where the generation of nonthermal particle spectra in magnetically dominated pair plasma through driven turbulence is investigated \citep[following ][]{zhdankin+17,groselj+24}.  
Here, we again employ the TRISTAN-MP v2 \citep{tristanv2} PIC code to study particle acceleration in turbulence.

We set up a computational domain of size $L^3$ and fill it with plasma of a uniform number density $n_0$ and temperature $k_B T/m_e c^2 = 10^{-2}$, where $k_B$ is the Boltzmann constant. For simplicity, we adopt electron-positron composition for the background plasma.
We do not expect this assumption to change the conclusions for the confinement and re-acceleration of test protons, as long as the total magnetization parameter, $\sigma=B^2/(4\pi (n_e m_p c^2 + n_e m_e c^2)$, is the same.
A mean guide field $B_0$ is introduced in the z-direction, with a magnitude set by $\sigma$, and a continuous turbulent state is established with a driving external current \citep{tenbarge+14}, thereby exciting Alfvènic perturbations on $L$ scales.
This setup mimics that of a “Langevin antenna”, where the frequency and decorrelation rate are set as $\omega_0 = 0.9 (2 \pi v_A / L)$, and $\gamma_0 = 0.5 \omega_0$, respectively, with $v_A = c \sqrt{\sigma/(1 + \sigma)}$.
The total amplitude of the fluctuations, is set as $\delta B=B_0$.
We consider periodic boundary conditions in all directions, and add a population of test protons, with a number density ratio of $\sim$1/100, injected with a flat spectrum until $\gamma_{\rm inj}=20$. 
Injected particles are allowed to deposit current in the box.
The plasma skin depth, $d_e$, is resolved with 2 cells and the box has a size $L^3 = (1920 d_e)^3$. 

Initial magnetic fluctuations are specified as follows, $\delta B_x = \sum_{h,i} \beta_{hi}i \sin(k_i x + \phi_{hi}) \cos(k_i y + \psi_{hi})$ and $\delta B_y = \sum_{h,i} \beta_{hi}h \cos(k_h x + \phi_{hi}) \sin(k_i y + \psi_{hi})$ \citep{comisso+18}, where the mode numbers are $h$ and $i$ with $h,i \in [1,..,\mathcal{H}]$, and $k_i = 2 \pi i/L$, and $k_h = 2 \pi h/L$. $\psi_{hi}$ and $\phi_{hi}$ are random phases.
In this setup, $\mathcal{H} = 8$ defines the energy-carrying scale $l$, such that $l = 2\pi/k_\mathcal{H} = L/\mathcal{H}$. 
The density fluctuations in the simulation at $t=L/c$ are plotted in Fig~\ref{fig:density}.

\subsection{Confinement}
Fig~\ref{fig:tnu-tesc} summarizes the interplay of cooling and confinement in NGC~1068 corona, as discussed in the main text. Fig~\ref{fig:confinement} shows the mean free path of injected particles from a proton burst in the turbulent box from Fig~\ref{fig:density}.

 \end{document}